\documentclass[12pt]{article}
\usepackage{amsfonts}
\usepackage{amssymb}
\usepackage{graphicx}
\usepackage{amsmath}
\usepackage{color}

\newtheorem{theorem}{Theorem}

\newtheorem{adefinition}[theorem]{Definition}

\newtheorem{aexample}[theorem]{Example}

\newtheorem{aproblem}[theorem]{Problem}

\newtheorem{acomment}[theorem]{Comment}

\newtheorem{aremark}[theorem]{Remark}

\numberwithin{equation}{section} \numberwithin{theorem}{section}

\makeatother \makeindex

\begin{document}

\title{On the Equilibrium State of a Small System with Random Matrix
Coupling to Its Environment}
\author{J.\ L.\ Lebowitz, \\
Department of Mathematics and Physics, Rutgers University, USA \and L.\
Pastur \\
Theoretical Department, Institute for Low Temperatures, Ukraine}
\date{}
\maketitle

\begin{abstract}
We consider a random matrix model of interaction between a small $n$-level
system, $S$, and its environment, a $N$-level heat reservoir, $R$. The
interaction between $S$ and $R$ is modeled by a tensor product of a fixed $%
n\times n$ matrix and a $N\times N$ hermitian Gaussian random matrix. We
show that under certain "macroscopicity" conditions on $R$, the reduced
density matrix of the system $\rho _{S}=\mathrm{Tr}_{R}\rho _{S\cup
R}^{(eq)} $, is given by $\rho _{S}^{(c)}\sim \exp {\{-\beta H_{S}\}}$,
where $H_{S}$ is the Hamiltonian of the isolated system. This holds for all
strengths of the interaction and thus gives some justification for using $%
\rho _{S}^{(c)}$ to describe some nano-systems, like biopolymers, in
equilibrium with their environment \cite{Se:12}. Our results extend those
obtained previously in \cite{Le-Pa:03,Le-Co:07} for a special two-level
system.
\end{abstract}



\section{Introduction}

The properties of a system $S$, in contact with a thermal reservoir $R$, is
an old yet perennial problem of statistical mechanics. Writing the
Hamiltonian of the composite system $S\cup R$ as
\begin{equation}
H_{S\cup R}=H_{S}\otimes \mathbf{1}_{R}+\mathbf{1}_{S}\otimes H_{R}+V_{SR}
\label{Htot}
\end{equation}%
where $H_{S}$ and $H_{R}$ are the Hamiltonians of the system and of the
reservoir and $V_{SR}$ is the interaction between them, the canonical Gibbs
density matrix of $S\cup R$ is given by
\begin{equation}
\rho _{S\cup R}^{(c)}=\exp \{-\beta H_{S\cup R}\}/Z_{S\cup R}.  \label{Gtot}
\end{equation}%
The corresponding reduced density matrix of the system is
\begin{equation}
\rho _{S}=\mathrm{Tr}_{R}\;\rho _{S\cup R}^{(c)}=\exp \{-\beta \widetilde{H}%
_{S}\}/\widetilde{Z}_{S}.  \label{red}
\end{equation}%
Here $\widetilde{H}_{S}$ is the "effective" Hamiltonian of $S$, which will
have the form
\begin{equation}
\widetilde{H}_{S}=H_{S}+\widetilde{V}_{S}  \label{Heff}
\end{equation}%
and $\widetilde{V}_{S}$ will in general depends on $\beta $, $H_{S},\ H_{R}$
and $V_{SR}$ unless $V_{SR}$ is "negligible" and $\rho _{S}$ can be replaced
by
\begin{equation}
\rho _{S}^{(c)}=\exp \{-\beta H_{S}\}/Z_{S}.  \label{GD}
\end{equation}%
The use of (\ref{GD}) for the density matrix of $S$ may be appropriate even
when $V_{SR}$ is not small, if the system $S$ is macroscopic, e.g., a system
in a large box $\Lambda ,$ and the interaction $V_{SR}$ takes place only
along the boundary $\partial \Lambda $. Then the calculations of the
properties of the system far from the boundary are approximately independent
of the interaction $V_{SR}$, becoming rigorously so when $\Lambda
\rightarrow \infty $ (there are exceptions when the system is at a first
order phase transition). Our concern here is however with the case when $S$
is small so the above considerations do not apply.

Recent technological advances, making it possible to create and manipulate
meso- and nanosystems, including biopolymers, colloidal particles, etc. have
brought the problem of a micro-system in contact with a reservoir to the
fore. Likewise, in recent studies of the foundation of quantum statistical
mechanics \cite{Ge-Co:05,Go-Le:06,Go-Le:10,Li-Co:10,Na-Hu:14,Po-Co:06},
where the approach to equilibrium is related to the system-reservoir
entanglement the notion of an equilibrium state for non macroscopic systems
also plays an important role. In all such cases the nature of the
interaction $V_{SR}$ between the system and the reservoir is clearly
important and $\rho _{S}$ is not necessarily of the Gibbs form (\ref{GD}). A
specific example of this importance is the collapse transition in polymers,
which depends strongly on the nature of solvent not just on its temperature
\cite{Hu:05}.

We note here that the distinction between the cases (\ref{red}) and (\ref{GD}%
) was made clearly by Jarzynski \cite{Ja:04} who showed that his equality
between the work done on the system by changing a parameter in $H_{S}$ from $%
A$ to $B$ is given by the difference of the free energies $\widetilde{F}_{S}$
\ given by $\log \widetilde{Z}_{S}$ in (\ref{red}) evaluated at the values
of the parameter $A$ and $B$. This means that when $V_{SR}$ is not
negligible then this difference will depend on $H_{R}$ and $V_{SR}$ and not
just on the temperature $\beta ^{-1}$ of the environment. This distinction
is sometimes blurred in the literature both experimental and theoretical
where various aspects of this problem are considered. Many of them go under
the name of stochastic thermodynamics, where the equilibrium state of a
nano-system in contact with an environment at temperature $\beta ^{-1}$ is
sometimes implicitly assumed to be described by the distribution (\ref{GD}),
see, e.g. reviews \cite{Ja:11,Ri:08,Se:12} and references therein. In fact
however these studies do not consider the whole microstate of the small
system. Instead, one argues that the reservoir degrees of freedom as well as
some "internal" degrees of freedom of the small system change very rapidly
compared to those observed, which we denote by $X$. They can therefore be
assumed to be in thermal equilibrium and their effect on the evolution of $X$
is then just described via some stochastic term (noise). This means in
particular that the only effect of the reservoir on the equilibrium state of
the $X$ variables is to specify its temperature, similar to the situation
for macroscopic systems. This seems reasonable in some cases (but not in
situations like those in the example of the polymer described earlier). One
may think then of the interaction term as fluctuating so rapidly that its
detailed nature "washes out" as far as its effect on the equilibrium
properties, or even the time evolution, of its slow variables are concerned.

In this paper we consider a simple model of such a situation. We do this by
making the interaction between the system and reservoir random. We obtain
the Gibbs distribution for an arbitrary finite-level quantum system $S$
using the frameworks of equilibrium statistical mechanics and assuming that
the reservoir has certain macroscopic properties.

Note that in this situation we can start, instead of (\ref{Gtot}), which
itself requires some justification \cite{Ja:04,Ja:11}, with the microcanonical
distribution for the composite system at a fixed interval $\Delta $ around a
macroscopic energy $E_{S\cup R}$, i.e., with
\begin{equation}
\rho _{S\cup R}(E_{S\cup R})=\chi _{\Delta }(H_{S\cup R})/\Omega _{S\cup
R}(E_{S\cup R}),  \label{mic}
\end{equation}%
where $\chi _{\Delta }$ is the indicator of the energy interval%
\begin{equation}
\Delta =(E_{S\cup R}-\delta /2,E_{S\cup R}+\delta /2),\ \delta <<E_{S\cup R}.
\label{eshell}
\end{equation}%
and
\begin{equation}
\Omega _{S\cup R}(E_{S\cup R})=\mathrm{Tr}_{S\cup R}\ \chi _{\Delta
}(H_{S\cup R}).  \label{micn}
\end{equation}%
We then obtain $\rho _{S}$ of (\ref{Heff}) with the inverse temperature
given by \cite{La-Li:80}
\begin{equation}
\beta =\frac{\partial \mathcal{S}_{R}}{\partial E_{R}}  \label{beta}
\end{equation}%
where
\begin{equation}
\mathcal{S}_{R}=\log \Omega _{R}(E_{R})  \label{SR}
\end{equation}%
is the entropy of the reservoir and $E_{R}$ is its energy (since for a
reservoir which is very large compared to the size of system we have $%
S_{R}\simeq S_{S\cup R}$ and $E_{R}\simeq E_{S\cup R},$).

The rest of the paper is organized as follows. The model is described in
Section 2, the results and discussion are given in Section 3. Sections 4 and
5 contain the proofs. The model is an extension of that introduced and
studied in our works \cite{Le-Pa:03,Le-Co:07}, where a simple case of a
2-level $S$ was considered both in the equilibrium and non-equilibrium
setting. Here we restrict ourselves to the equilibrium and leave dynamic
considerations for a future work.

According to the above, the proofs are given for the microcanonical
distribution (\ref{mic}) -- (\ref{SR}) of the composite system. The results
for the canonical distribution (\ref{Gtot}) of the composite prove to be a
simple corollary of those for the microcanonical distribution and require a
version of a standard (in statistical mechanics) saddle point argument.

Note also that large random matrices have been widely used to model a
variety of complex quantum systems, including heavy nuclei and atoms,
mesoscopic particles, quantum networks, graphs etc., i.e., not necessarily
macroscopic, systems \cite{Ak-Co:11,Gu-Co:98,Pa-Sh:11}. Thus, our setting
could model a qubit or a microcluster in a mesoscopic (or even nanoscopic)
particle, a quantum dot, small quantum network, etc.

\section{Model}

We describe now the model and quantities that will be considered. Let $H_{R}$
be a $N\times N$ hermitian matrix, $\{E_{j}\}_{j=1}^{N}$ \ be its
eigenvalues and
\begin{equation}
\nu _{N}(E)=N^{-1}\sum_{j=1}^{N}\delta (E-E_{j}),\qquad \int_{-\infty
}^{\infty }\nu _{N}(E)dE=1.  \label{non}
\end{equation}%
be its density of states normalized to unity. We assume that $\nu _{N}$
converges as $N\rightarrow \infty $ to a continuous density $\nu $ in the
sense that for any continuous and bounded function $f$ we have:
\begin{equation}
\lim_{N\rightarrow \infty }\int_{-\infty }^{\infty }f(E)\nu
_{N}(E)dE=\int_{-\infty }^{\infty }f(E)\nu (E)dE,\;\int_{-\infty }^{\infty
}\nu (E)dE=1.  \label{no}
\end{equation}%

Let $W_{R}$ be a random hermitian matrix, distributed according to the
Gaussian unitary invariant law given by probability density
\begin{equation}
\mathcal{Z}_{R}^{-1}\exp \left\{ -\mathrm{Tr\;}W_{R}^{2}/2\right\} ,
\label{GUE}
\end{equation}%
where $\mathcal{Z}_{R}$ is the normalization constant. In other words, we
assume that the entries of $W_{R}=\{W_{jk}\}_{j,k=1}^{N},\;W_{kj}=\overline{%
W_{jk}}$ are complex and independent for $1\leq j\leq k\leq N$ Gaussian
random variables such that
\begin{equation}
\mathbf{E}\{W_{jk}\}=\mathbf{E}\{W_{jk}^{2}\}=0,\;\mathbf{E}%
\{|W_{jk}|^{2}\}=(1+\delta _{jk}).  \label{GOE}
\end{equation}%
This is known as the Gaussian Unitary Ensemble (see e.g. \cite{Pa-Sh:11},
Section 1.1).

Let also $H_{S}$ and $\Sigma _{S}$ be arbitrary $n\times n$ hermitian
matrices. We define the Hamiltonian of our composite system $S\cup R$ as a
random $nN\times nN$ matrix (cf. (\ref{Htot}))
\begin{equation}
H_{S\cup R}=H_{S}\otimes \mathbf{1}_{N}+\mathbf{1}_{n}\otimes H_{R}+\Sigma
_{S}\otimes W_{R}/N^{1/2},  \label{Ham}
\end{equation}%
where $\mathbf{1}_{l}$ is the unit $l\times l$ matrix.

It should be noted that our results are valid not only for the special
Gaussian distributed interaction (\ref{GUE}) -- (\ref{GOE}), but for any
real symmetric or hermitian random matrix in (\ref{Ham}), whose entries $%
W_{jk},\;1\leq j\leq k\leq N$ are independent and satisfy (\ref{GOE}).
However, in this case the techniques are more involved requiring an
extension of those of random matrix theory for so called Wigner matrices
(see \cite{Pa-Sh:11}, Section 18.3).

In addition, our results are also valid in the case, where $H_{R}$ is random
and independent of $W_{R}$ for all $N$. In this case we have to assume that
the sequence $\{H_{R}\}$ of random matrices is defined for all $N\rightarrow
\infty $ on the same probability space as the sequence $\{W_{R}\}$, is
independent of $\{W_{R}\}$ and satisfies (\ref{no}) with probability 1.

Having $H_{R}$ and $W_{R}$ random, we can view the second term in (\ref{Ham}%
) as the Hamiltonian of a "typical" $N$-level reservoir and the third term
as a "typical" interaction between the system and its reservoir. It is worth
mentioning that the notion of typicality has recently been used in the
studies of the foundations of quantum statistical mechanics, including the
form of reduced density matrix in equilibrium \cite%
{Go-Le:06,Go-Le:10,Li-Co:10,Na-Hu:14,Po-Co:06}. In addition, the randomness
(frozen disorder) is a basic ingredient of the theory of disordered systems.
Its successful and efficient use is justified by establishing the
selfaveraging property of the corresponding results, i.e., their validity
for the overwhelming majority of realizations of randomness for
macroscopically large systems, see e.g. \cite{LGP}.

In our case the selfaveraging property is valid in the limit $N\rightarrow
\infty $ and is given by Result \textbf{I} below. This suggests that in our
model the $N\rightarrow \infty $ limit plays the role of the macroscopic
limits in statistical mechanics and condensed matter theory. Note however that in
statistical mechanics the density of states of a macroscopic reservoir is
multiplicative in its volume and/or in its number of degrees of freedom.
This and the macroscopicity of the system lead to the Gibbs form (\ref{GD})
of its reduced density matrix, if the system-reservoir interaction is of
short range and confined to the boundary of $S$ \cite{La-Li:80}.

If, however, the system is small, then, as is already noted, its reduced
density matrix is not (\ref{GD}) in general even if the reservoir is
macroscopically large. This is well understood in statistical mechanics and
in principle in the stochastic thermodynamics community \cite{Ja:04,Ja:11}. This is
especially in the frameworks of dynamical approach where the problem was
first studied by Bogolyubov \cite{Bo:45} for a classical system
consisting of a harmonic oscillator interacting linearly with the
macroscopic reservoir of harmonic oscillators, and then in a number of
interesting and rather general quantum models with macroscopic many-body
reservoirs, see, e,g., \cite{At-Co:07,Ba-Co:00,De-Ku:11}. Taking into
account that according to (\ref{non}) and (\ref{no}) the density of states $%
N\nu _{N}$ of our reservoir is asymptotically additive (but not
multiplicative) in $N$, we conclude that the Gibbs form of the reduced
density matrix seems to be even less likely in our case than in the
spin-boson model and that one needs additional conditions and procedures to
obtain the Gibbs distribution (\ref{GD}). We will discuss such conditions
below. Here we only mention that an asymptotically additive in volume
density of states is the case in the one-body approximation of solid state
physics.


\smallskip

According to \cite{Le-Pa:03} the Gibbs distribution for the two-level ($n=2$%
) version of our model can be obtained if we assume that the reservoir
consists of large number $1<<J<<N$ of independent or weakly dependent parts,
more precisely, that the normalized density of states $\nu _{N}$ of the
reservoir has a special "quasi-multiplicative" form (\ref{nuj}). One can
view this assumption as an effort to obtain the multiplicativity of the
density of states of the reservoir on a scale which is intermediate
(mesoscopic) between the microscopic and macroscopic scales. It is shown
below that the same assumption on $\nu _{N}$ allows one to obtain the Gibbs
distribution (\ref{GD}) for an arbitrary finite dimensional $S$.

\section{Results}

We are interested in the asymptotic form as $N\rightarrow \infty $ of the
\textit{reduced density matrix}%
\begin{equation}
\rho _{S}^{(N)}=\frac{N^{-1}\mathrm{Tr}_{S}\;\chi _{\Delta }(H_{S\cup R})}{%
N{}^{-1}\mathrm{Tr}\;\chi _{\Delta }(H_{S\cup R})}  \label{mer}
\end{equation}%
corresponding to the microcanonical distribution (\ref{mic}) -- (\ref{micn})
of our model composite system (\ref{Ham}).

\smallskip Our first result is

\medskip \textit{\textbf{I}. There exists a positive definite, trace one,
non-random matrix }$\rho_S$\textit{\ such that for any fixed }$\Delta $
\textit{of} (\ref{eshell}) \textit{we have with probability 1}%
\begin{equation}
\lim_{N\rightarrow \infty }\rho _{S}^{(N)}=\rho_S.  \label{rsr}
\end{equation}%
The result is proved in Section 4.

To describe the limiting density matrix $\rho _{S}$, we start from the
relation
\begin{equation}
\chi _{\Delta }(H_{S\cup R})=\mathcal{E}_{H_{S\cup R}}(\Delta ),  \label{Ehs}
\end{equation}%
where $\mathcal{E}_{H_{S\cup R}}$ is the resolution of identity (spectral
projection) of the hermitian operator $H_{S\cup R}$ (\ref{Ham})
corresponding to the spectral interval $\Delta $ of (\ref{eshell}). Denoting%
\begin{equation}
e_{S}^{(N)}(d\lambda )=N^{-1}\mathrm{Tr}_{R}\;\mathcal{E}_{H_{S\cup
R}}(d\lambda ),  \label{ehs}
\end{equation}%
we can write, using (\ref{eshell}), (\ref{Ehs}) and (\ref{mer}),%
\begin{equation}
\rho _{S}^{(N)}=\frac{e_{S}^{(N)}(\Delta )}{\mathrm{Tr}_{S}\;e_{S}^{(N)}(%
\Delta )}.  \label{mer1}
\end{equation}%
Thus, we have to find the limit
\begin{equation}
e_{S}(\Delta )=\lim_{N\rightarrow \infty }e_{S}^{(N)}(\Delta )  \label{lehs}
\end{equation}%
of the $n\times n$ random hermitian matrix $e_{S}^{(N)}(\Delta )$ and then
we can write (\ref{rsr}) as
\begin{equation}
\rho _{S}=\frac{e_{S}(\Delta )}{\mathrm{Tr}_{S}\;e_{S}(\Delta )}  \label{rbe}
\end{equation}%
To find $e_S(\Delta )$ of (\ref{lehs}) we will use the resolvent%
\begin{equation}
G(z):=(H_{S\cup R}-z)^{-1},\;\Im z\neq 0  \label{G}
\end{equation}%
of $H_{S\cup R}$. Indeed, we have, by the spectral theorem for hermitian
matrices,
\begin{equation}
G(z)=\int \frac{\mathcal{E}_{H_{S\cup R}}(d\lambda )}{\lambda -z},\;\Im
z\neq 0,  \label{Gz}
\end{equation}%
and we write here and below integrals without limits for those over the
whole real axis. It follows then from (\ref{ehs}) and (\ref{Gz}) that%
\begin{equation}
g_{S}^{(N)}(z):=N^{-1}\mathrm{Tr}_{R}\;G(z)=\int \frac{e_{S}^{(N)}(d\lambda )%
}{\lambda -z},\;\Im z\neq 0.  \label{gz}
\end{equation}%
Note that $e_{S}$ is the matrix valued measure assuming values in $n\times n$
positive definite matrices and is uniformly bounded in $N$, since for any $%
\Delta \subset \mathbb{R}$ the matrix $\mathcal{E}_{H_{S\cup R}}(\Delta )$,
being an orthogonal projection in $nN$ space $\mathcal{H}_{S}\otimes
\mathcal{H}_{R}$, is of norm one and according to (\ref{ehs})
\begin{equation*}
||e_{S}^{(N)}(\Delta )||\leq n.
\end{equation*}%
Here $||A||$ is the standard matrix norm of a matrix $A$ and we used the
inequality%
\begin{equation*}
||\mathrm{Tr}_{R}\;A||\leq nN||A||,
\end{equation*}%
valid for any hermitian matrix $A$ in $\mathcal{H}_{S}\otimes \mathcal{H}%
_{R} $.

Recall now that for any non-negative finite measure $m$ on the real axis one
can define its \textit{Stieltjes transform}%
\begin{equation}
s(z)=\int \frac{m(d\lambda )}{\lambda -z},\;\Im z\neq 0,  \label{ST}
\end{equation}%
which is analytic for $\Im z\neq 0$ and such that
\begin{equation}
\Im s(z)\ \Im z>0,\;\Im z\neq 0.  \label{imim}
\end{equation}%
The correspondence between non-negative measures and their Stieltjes
transforms is one-to-one, in particular%
\begin{equation}
m(\Delta )=\lim_{\delta \rightarrow 0^{+}}\frac{1}{\pi }\int_{\Delta }\Im
s(\lambda +i\delta )d\lambda .  \label{SP}
\end{equation}%
Besides, the correspondence is continuous with respect to the weak
convergence of measures (see e.g. (\ref{no})) and the uniform convergence of
their Stieltjes transforms on a compact set of $\mathbb{C}\setminus \mathbb{R%
}$ (see e.g. \cite{Pa-Sh:11}, Proposition 2.1.2).

It is easy to extend (\ref{ST}) -- (\ref{SP}) to the matrix valued positive
definite and bounded measures and their matrix valued Stieltjes transforms,
whose examples are (\ref{ehs}) and (\ref{gz}) respectively. Thus, to prove (%
\ref{rbe}) it suffices to prove that for a compact set of $\mathbb{C}%
\setminus \mathbb{R}$ the matrix valued functions (\ref{gz}) converge with
probability 1 to a non-random limit on a compact set in $\mathbb{C}\setminus
\mathbb{R}$.

\medskip

Correspondingly, we prove in Sections 4 our second result

\medskip \textit{\textbf{II. }Let }$e_{S}^{(N)}(\Delta ),\;g_{S}^{(N)},\ \nu
,\;H_{S}$\textit{\ and }$\Sigma _{S}$ \textit{be defined by (\ref{ehs}), (%
\ref{Gz}) -- (\ref{gz}), (\ref{non}) -- (\ref{no}) and (\ref{Ham}). Then}$:$

(i)\textit{\ the limit }$e_{S}(\Delta )$ of \textit{(\ref{lehs}) exists with
probability 1, hence (\ref{rsr}) and (\ref{rbe}) hold with the same
probability;}

(ii) \textit{if }$f_{S}$\textit{\ is the matrix Stieltjes transform of }$%
e_{S}$\textit{, i.e., }%
\begin{equation}
e_{S}(\Delta )=\lim_{\delta \rightarrow 0^{+}}\frac{1}{\pi }\int_{\Delta
}\Im f_{S}(\lambda +i\delta )d\lambda  \label{eF}
\end{equation}%
\textit{then} $f_S$ \textit{is a unique solution of the matrix equation}%
\begin{equation}
f_{S}(z)=\int \left( E+H_{S}-z-\Sigma_{S} f_{S}(z)\Sigma _{S}\right)
^{-1}\nu (E)dE  \label{MPm}
\end{equation}%
\textit{in the class of matrix valued functions analytic for }$\Im z\neq 0$%
\textit{\ and such that (cf. (\ref{imim}))}

\begin{equation}
(f_{S}(z)-f_{S}^{\ast }(z))/(z-\overline{z})>0,\;\Im z\neq 0,  \label{nevm}
\end{equation}%
\textit{and for any hermitian matrix }$A$\textit{\ we write }$A>0$\textit{\
if }$A$\textit{\ is positive definite. }

\medskip \textbf{Remark}. The case $n=1$ of the above assertion corresponds
to a particular case of the deformed Gaussian Ensembles of random matrix
theory (see \cite{Pa-Sh:11}, Section 2, Theorem 2.2.1 in particular). The
case $n=2$, $H_{S}=s\sigma _{z},\ s>0$, and $\Sigma _{S}=\sigma _{z}$, where
$\sigma _{x}$ and $\sigma _{z}$ are the corresponding Pauli matrices, was
considered in \cite{Le-Pa:03,Le-Co:07}, while studying a random matrix model
of quantum relaxation dynamics.

\medskip The obtained limiting reduced density matrix $\rho _{S}$ (see (\ref%
{rsr}) and (\ref{rbe})) is generally not of the Gibbs form (\ref{GD}). We
thus need additional assumptions on the structure of reservoir in order to
obtain the Gibbs distribution in the frameworks of our model. In our
previous work \cite{Le-Pa:03}, where the case $n=2$ was considered, it was
assumed that the reservoir consists of a large number $J\rightarrow \infty $
of practically non-interacting "macroscopically infinitesimal" but also
sufficiently large parts (i.e., a kind of "coarse grained" structure of
reservoir). This can be implemented by writing the density of states (\ref%
{no}) of reservoir as the convolution of $J$ copies of a certain density $q$:%
\begin{equation}
\nu :=\nu _{J}=q^{\ast J}.  \label{nuj}
\end{equation}%
The fact that we are going to consider the asymptotic regime $J\rightarrow
\infty $ after the limit $N\rightarrow \infty $ can be interpreted as a
formalization of the inequality determining our intermediate scale%
\begin{equation}
n<<J<<N,  \label{jineq}
\end{equation}%
or, denoting $n_{q}=N/J$ the parameter characterizing the "size" of
infinitesimal parts, as the condition $n_{q}>>1$.

A simple example of the above is the Gaussian density
\begin{equation}
q(\varepsilon )=\frac{1}{(2\pi a^{2})^{1/2}}\exp \left\{ -\frac{(\varepsilon
-\varepsilon _{0})^{2}}{2a^{2}}\right\} ,  \label{ga0}
\end{equation}%
where $\varepsilon _{0}$ is assumed to be of the order of magnitude of the
characteristic energies of "macroscopically infinitesimal" parts of $R$, and
$a$ is their energy spread. In this case the density of states (\ref{nuj})
of the reservoir is also Gaussian
\begin{equation}
\nu_{J} (E)=\frac{1}{(2\pi Ja^{2})^{1/2}}\exp \left\{ -\frac{(E-J\varepsilon
_{0})^{2}}{2Ja^{2}}\right\} .  \label{gauss}
\end{equation}%
The formula makes explicit one more property of our "macroscopic" reservoir:
its characteristic energies are of order of $J\varepsilon _{0}$ with the
spread $(Ja^{2})^{1/2}$, while those of $H_{S}$ and $\Sigma _{S}$ are
independent of $J$, hence, much smaller.

The Gaussian "ansatz" (\ref{ga0}) -- (\ref{gauss}) was used in \cite%
{Le-Pa:03,Le-Co:07}. In this paper we study the equilibrium properties of (.%
\ref{Ham}) with an arbitrary finite $n$ and $n\times n$ hermitian $H_{S}$
and $\Sigma _{S}$ in (\ref{Ham}). As already noted, our results are valid
for a wide and natural class of "macroscopically infinitesimal" part
distributions $q$ in (\ref{nuj}), see Section 5. In particular, one can
mention the densities
\begin{equation}
q_{d}=q_{1}^{\ast d},\;q_{1}(\varepsilon )=\Big(\pi \sqrt{\varepsilon
(\varepsilon _{0}-\varepsilon )}\Big)^{-1},\;0\leq \varepsilon \leq
\varepsilon _{0},  \label{qd}
\end{equation}%
where $q_{1}$ is the density of states of the one dimensional harmonic chain
and $q_{d}$ is the density of states of the $d$-dimensional harmonic cubic
lattice. Different choices of $q$ can model different types of reservoirs.
For carrying out the proofs we shall require two conditions on $q$:

\medskip

\smallskip (i) $q$ decays superexponentially at $-\infty $;

\smallskip

(ii) there exists $1<a\leq 2$ such that%
\begin{equation}
\int q^{a}(\varepsilon )d\varepsilon <\infty .  \label{phidec}
\end{equation}%
The conditions seem fairly natural. Indeed, condition (i) requires a rather
"thin" if any spectrum of negative energies of large absolute value, thereby it
is closely related to the stability of the reservoir. Furthermore,  since (%
\ref{phidec}) is valid by definition for $a=1$ (recall that $q$ is a
probability density), condition (ii) requires a certain regularity of the
density of states, also usually assumed in statistical mechanics. Note also
that if  $q$ is zero for large negative energies, then one can use the
polynomial decay of the Fourier transform of $q$ as another condition of its
regularity.

Here are simple examples of the above. The first is the Gaussian density (%
\ref{ga0}), for which the validity of the Gibbs distribution in the
asymptotic regime (\ref{jineq}) for $n=2$ was proved in \cite{Le-Pa:03}.
Here $a$ is any real number of $(1,2]$. The second example is the
exponential density $q(\varepsilon )=e^{-\varepsilon /\varepsilon
_{0}}\varepsilon _{0}^{-1}$, where again $a\in (1,2]$. The third example is
the density of states (\ref{qd}) of the simple cubic $d$-dimensional
crystal. Here $a\in (1,2)$ for $d=1$ and $a\in (1,2]$ for $d\geq 2$.

Viewing the density of states (\ref{nuj}) as the partition function of the
microcanonical ensemble of an ideal gas of $J$ particles having each the
partition function $q$ one can introduce an analog of the entropy per
particle
\begin{equation}
s(\varepsilon ):=\lim_{J\rightarrow \infty }J^{-1}\log \nu _{J}(J\varepsilon
).  \label{entg}
\end{equation}%
The existence of the limit, its continuity and convexity can be proved by
now standard argument of statistical mechanics (see e.g. \cite%
{Ru:69,Mi-Po:67}).

One can also introduce an analog of the inverse temperature (cf. (\ref{beta}%
))
\begin{equation}
\beta (\varepsilon )=s^{\prime }(\varepsilon ).  \label{bsp}
\end{equation}%
The corresponding quantities for the canonical ensemble of "macroscopically
infinitesimal" parts of our reservoir are the analogs of the partition
function and the free energy per particle
\begin{equation}
Z_{J}(\beta ):=\int e^{-\beta E}\nu _{J}(E)dE=\psi ^{J}(\beta ),\;f(\beta
):=-(\beta J)^{-1}\log Z_{J}(\beta )=-\beta ^{-1}\log \psi (\beta ),
\label{zb}
\end{equation}%
where%
\begin{equation}
\psi (\beta )=\int e^{-\beta \varepsilon }q(\varepsilon )d\varepsilon .
\label{psi}
\end{equation}%
In writing the partition function (\ref{zb}) we took into account that the
Laplace transform of the $J$-fold convolution $\nu _{J}$ of (\ref{nuj}) is $%
J $th power of the Laplace transform of $q$. Condition (i) above ensures
that $\psi $ is well defined for all non-negative $\beta $.

We have also from (\ref{entg}) and (\ref{zb}) -- (\ref{psi}).%
\begin{equation}
\beta f(\beta )=\min_{\varepsilon }\{\varepsilon \beta -s(\varepsilon )\}
\label{fleg}
\end{equation}%
and%
\begin{equation}
s(\varepsilon )=\min_{\beta }h_{\varepsilon }(\beta ),\;h_{\varepsilon
}(\beta )=\varepsilon \beta -f(\beta )  \label{sleg}
\end{equation}%
Note that $\beta f$ and $s$ are convex. The convexity of $\beta f$ is
immediate, since (\ref{zb}) and (\ref{psi}) imply
\begin{equation}
(\beta f)^{\prime \prime }=-\left( \frac{\psi ^{\prime \prime }}{\psi }-%
\frac{\psi ^{\prime 2}}{\psi ^{2}}\right) =-d,\;d:=\int (\varepsilon
-\varepsilon _{0})^{2}Q(\varepsilon )d\varepsilon >0  \label{fconv}
\end{equation}%
where%
\begin{equation}
Q(\varepsilon )=e^{-\beta \varepsilon }q(\varepsilon )\left( \int e^{-\beta
\varepsilon ^{\prime }}q(\varepsilon ^{\prime })d\varepsilon ^{\prime
}\right) ^{-1}\geq 0  \label{qe}
\end{equation}%
and $\varepsilon _{0}$\ is the first moment of $Q$. As for the convexity of $%
s$, it follows from the proof of (\ref{entg}) \cite{Ru:69,Mi-Po:67}.

We have also from (\ref{zb}) -- (\ref{sleg})
\begin{equation}
h_{\varepsilon }(0)=0,\;h_{\varepsilon }^{\prime }(0)=\varepsilon -\overline{%
\varepsilon },  \label{hpo}
\end{equation}%
where%
\begin{equation*}
\overline{\varepsilon }=\int \varepsilon q(\varepsilon )d\varepsilon
\end{equation*}%
is the mean energy of the macroscopically infinitesimal components of the
reservoir. We will assume that%
\begin{equation}
\varepsilon >\overline{\varepsilon },  \label{eeb}
\end{equation}%
since then the convexity of $h_{\varepsilon }$ and (\ref{hpo}) imply that
the minimum $s(\varepsilon )$ of $h_{\varepsilon }$ in (\ref{sleg}) is
positive, hence our temperature (\ref{bsp}) is positive as well
\begin{equation}
\beta (\varepsilon )>0,  \label{bepo}
\end{equation}%
although negative temperatures can also be considered in the frameworks of
our model.

It turns out, however, that the above standard formulas of statistical
mechanics of macroscopic systems are not accurate enough to obtain the Gibbs
form (\ref{GD}) of the reduced density matrix (\ref{eF}) -- (\ref{MPm}) of $%
S $ in our model (\ref{nuj}) of reservoir. This is because of the
"logarithmic accuracy" of the formulas, see e.g. (\ref{entg}), giving the
large $J$ leading term of $\log \nu _{J}(J\varepsilon )$, while we will need
below the more accurate asymptotics formula:
\begin{equation}  \label{df}
\nu _{J}(J\varepsilon )=\mu _{J}(\varepsilon )(1+o(1)),\;J\rightarrow \infty
,\;\mu _{J}(\varepsilon ):=(2\pi Js^{\prime \prime }(\varepsilon
))^{-1/2}e^{-Js(\varepsilon )}.
\end{equation}%
This formula can be viewed as the Darwin-Fawler version of the equivalence
of the microcanonical and canonical ensembles in statistical mechanics (see
e.g. \cite{Hu:87}).

In fact, we do not need to know that $\mu _{J}(\varepsilon )$ is the
asymptotic (\ref{df}) of $\nu _{J}(J\varepsilon )$ to prove our third result
below, i.e., the validity of the Gibbs form of the energy distribution of
the system in our model (\ref{nuj}) of the reservoir. Instead, we will just
use $\mu _{J}$ of (\ref{df}) as an ansatz. Since, however, our proof is
applicable to all finite $n\geq 1$ and all $n\times n$ hermitian $H_{S}$ and
$\Sigma _{S}$ in (\ref{Ham}) and since the case $n=1$, and $H_{S}=\Sigma
_{S}=0$ corresponds to the reservoir itself, thus, in this case $\gamma
_{J}(E)=\nu _{J}(E)$, we obtain the asymptotic formula (\ref{df}) as the
simplest case $n=1$ of (\ref{gamu}).

\textit{\textbf{III.} Under the conditions of validity of Results I and II\
above and for the density of states of reservoir given by (\ref{nuj}) with }$%
q$ \textit{satisfying conditions (i) and (ii), (\ref{phi}) and (\ref{phidec}%
) in particular, we have:}

(i)\textit{\ the reduced density matrix of (\ref{rsr}) and (\ref{rbe}) has a
well defined limit as the width }$\delta $\textit{\ of the energy shell }$%
\Delta $ of \textit{\ (\ref{eshell})) tends to zero, i.e., }$\Delta
\rightarrow \{E\}$\textit{\ and }$J$\textit{\ is large enough:}%
\begin{equation}
\widehat{\rho }_{J}(E):=\lim_{\Delta \rightarrow \{E\}}\rho _{S}=\frac{%
\gamma _{J}(E)}{\mathrm{Tr}_{S}\gamma _{J}(E)},\;\gamma _{J}(E)=\lim_{\Delta
\rightarrow \{E\}}e_{S}(\Delta ),  \label{rgg}
\end{equation}%
\textit{where }$e_S(\Delta )$\textit{\ is given by }(\ref{lehs});

(ii) \textit{the limiting relation }%
\begin{equation}
\lim_{J\rightarrow \infty }\gamma _{J}(J\varepsilon )/\mu _{J}(\varepsilon
)=e^{-\beta H_{S}},  \label{gamu}
\end{equation}%
\textit{with }$\mu _{J}$\textit{\ given in (\ref{df}), hence the Gibbs form (%
\ref{GD}) of the limit (in view of (\ref{rgg}))}%
\begin{equation}
\lim_{J\rightarrow \infty }\widehat{\rho }_{J}(J\varepsilon )=\frac{%
e^{-\beta H_{S}}}{\mathrm{Tr}_{S}e^{-\beta H_{S}}},  \label{limgi}
\end{equation}%
\textit{in which }$\beta :=\beta (\varepsilon )$\textit{\ is defined by} (%
\ref{entg}) -- \textit{(\ref{bsp}) and (\ref{eeb}) -- (\ref{bepo}).}

\medskip The result is proved in Section 5.

\medskip \textbf{Remark.} The above result concerns the microcanonical
energy distribution (\ref{mic}) -- (\ref{micn}) of the composite system and
the corresponding reduced density matrix (\ref{mer}). The case of the
canonical distribution (\ref{Gtot}) of the whole system and the
corresponding reduced density matrix, i.e., the passage from (\ref{Heff}) to
(\ref{GD}) can be readily obtained from (\ref{rgg}) -- (\ref{limgi}).
Indeed, we have from (\ref{Gtot}), the spectral theorem for $H_{S\cup R}$, (%
\ref{ehs}) and (\ref{rgg}) that with probability 1:%
\begin{equation}
\lim_{N\rightarrow \infty }\frac{\mathrm{Tr}_{R}e^{-\beta H_{S\cup R}}}{%
\mathrm{Tr}_{S}\mathrm{Tr}_{R}e^{-\beta H_{S\cup R}}}=\lim_{N\rightarrow
\infty }\frac{\int e^{-\beta E}e_{S}(dE)}{\int e^{-\beta E}\mathrm{Tr}%
_{S}e_{S}(dE)}=\frac{\int e^{-\beta E}\gamma _{J}(E)dE}{\int e^{-\beta E}%
\mathrm{Tr}_{S}\gamma _{J}(E)dE}.  \label{cantu}
\end{equation}%
Using then (\ref{df}), (\ref{gamu}) and a simple saddle point argument for
integrals on the r.h.s. of (\ref{cantu}) yield%
\begin{equation}
\lim_{J\rightarrow \infty }\frac{\int e^{-J(\beta \varepsilon -s(\varepsilon
))}\gamma _{J}(J\varepsilon )\mu _{J}^{-1}(\varepsilon )d\varepsilon }{\int
e^{-\beta E}\mathrm{Tr}_{S}\gamma _{J}(J\varepsilon )\mu
_{J}^{-1}(\varepsilon )d\varepsilon }=\frac{e^{-\beta H_{S}}}{\mathrm{Tr}%
_{S}e^{-\beta H_{S}}},  \label{canj}
\end{equation}%
i.e., the Gibbs distribution again.

\section{ Selfaveraging Property and Limiting Reduced Density Matrix}

In this section we prove Results 1 and II above. We note that the
corresponding assertions as well as their proofs are generalizations of
those for the deformed semicircle law (DSCL) of random matrix theory, see
\cite{Pa-Sh:11}, Sections 2.2 and 18.3. Just as in the case of DSCL the
passage from the positive definite matrix measures (positive measures in the
case of DSCL) to their Stieltjes transforms (see (\ref{mer1}) and (\ref{gz}%
)) reduces the proof of (\ref{mer}) -- (\ref{rsr}) to that of (\ref{MPm}) --
(\ref{nevm}). The latter facts will follow if we prove that the variance of
all the entries of (\ref{gz}) vanish fast enough as $n\rightarrow \infty $
and that the expectation of (\ref{gz}) converges to a unique solution of (%
\ref{MPm}) as $n\rightarrow \infty $.

We will use the Greek indices varying from $1$ to $n$ to label the states of
the systems and the Latin indices varying from $1$ to $N$ to label the
states of reservoir.

Then we can write $n\times n$ and $nN\times nN$ matrices of (\ref{gz}) and (%
\ref{G}) as%
\begin{equation}
g_{S}^{(N)}(z)=\{g_{\alpha \beta }(z)\}_{\alpha ,\beta =1}^{n},\;\{G_{\alpha
j,\beta j}(z)\}_{\alpha ,\beta =1;j,k=1}^{n,N}  \label{gG}
\end{equation}%
and
\begin{equation}
g_{\alpha \beta }(z)=\frac{1}{N}\sum_{j=1}^{N}G_{\alpha j,\beta j}(z).
\label{gTG}
\end{equation}%
We will prove now the bound%
\begin{equation}
\mathbf{Var}\{g_{\alpha \beta }(z):=\mathbf{E}\{|g_{\alpha \beta
}(z)|^{2}\}-|\mathbf{E}\{g_{\alpha \beta }(z)\}|^{2}\leq C_{S}(z)/N^{2},
\label{varab}
\end{equation}%
and we write here and below $C_{S}(z)$ for quantities which do not depend on
$N$ and are finite for $\Im z\neq 0$.

To this end we view every $g_{\alpha \beta }$ as a function of the Gaussian
random variables $\{W_{jk}\}_{j,k=1}^{N}$ of (\ref{GUE}) -- (\ref{GOE}) and
use the Poincar\'{e} inequality (see \cite{Pa-Sh:11}, Proposition 2.1.6),
yielding%
\begin{equation}
\mathbf{Var}\{g_{\alpha \beta }(z)\}\leq \sum_{j,k=1}^{N}\mathbf{E}\left\{
\left\vert \frac{\partial g_{\alpha \beta }}{\partial W_{jk}}\right\vert
^{2}\right\} .  \label{vab}
\end{equation}%
The derivatives on the right can be found by using the resolvent identity
for $G$%
\begin{eqnarray*}
\frac{\partial g_{\alpha \beta }(z)}{\partial W_{jk}} &=&-\frac{1}{N^{3/2}}%
\sum_{\gamma ,\delta -1}^{n}(G_{\delta \beta }(z)G_{\alpha \gamma
}(z))_{jk}\Sigma _{\gamma \delta } \\
&\leq &\left( \frac{1}{N^{3}}\sum_{\gamma ,\delta -1}^{n}|(G_{\delta \beta
}(z)G_{\alpha \gamma }(z))_{jk}|^{2}\sum_{\gamma ,\delta -1}^{n}|\Sigma
_{\gamma \delta }|^{2}\right) ^{1/2},
\end{eqnarray*}%
where $G_{\alpha \beta }$ denotes the $N\times N$ matrix
\begin{equation*}
G_{\alpha \beta }(z)=\{G_{\alpha j,\beta k}(z)\}_{j ,k =1}^{N}.
\end{equation*}%
The $n\times n$ matrix $\Sigma _{S}=\{\Sigma _{\alpha \beta }\}_{\alpha
,\beta =1}^{n}$ is defined by (\ref{Ham}) and the second line results from
Schwarz's inequality. Taking into account the bounds%
\begin{equation}
||G(z)||\leq |\Im z|^{-1},\;|G_{\alpha j,\beta k}(z)|\leq ||G(z)||\leq |\Im
z|^{-1}  \label{Gbou}
\end{equation}%
valid for the resolvent of any hermitian matrix, it is easy to find an
analogous bound for the matrix $G_{\alpha \beta }(z)$:
\begin{equation}
||G_{\alpha \beta }(z)||\leq |\Im z|^{-1}.  \label{ngab}
\end{equation}%
The above relations imply for the r.h.s. of (\ref{vab})
\begin{equation*}
\mathbf{Var}\{g_{\alpha \beta }(z)\}\leq \frac{1}{N^{3}}\mathrm{Tr}%
_{S}\Sigma _{S}^{2}\sum_{\gamma ,\delta -1}^{n}\mathrm{Tr}_{R}(G_{\delta
\beta }G_{\alpha \gamma }G_{\alpha \gamma }^{\ast }G_{\delta \beta }^{\ast })
\end{equation*}%
and using now (\ref{ngab}) and the inequality $|\mathrm{Tr}A|\leq N||A||$
valid for any $N\times N$ matrix, we obtain%
\begin{equation*}
\mathbf{Var}\{g_{\alpha \beta }(z)\}\leq \frac{n^{2}}{N^{2}|\Im z|^{4}}%
\mathrm{Tr}_{S}\Sigma _{S}\Sigma _{S}^{\ast },
\end{equation*}%
i.e., the bound (\ref{varab}) with $C_{S}(z)=n^{2}\mathrm{Tr}_{S}\Sigma
_{S}^{2}/|\Im z|^{4}$.

Denote
\begin{equation}
f_{S}^{(N)}(z)=\{f_{\alpha \beta }(z)\}_{\alpha ,\beta =1}^{n},\;f_{\alpha
\beta }(z)=\mathbf{E}\{g_{\alpha \beta }(z)\}.  \label{fN}
\end{equation}%
We will now prove that for any compact set $K\subset \mathbb{C}\diagdown
\mathbb{R}$ there is a subsequence $\{f_{S}^{(N_{i})}\}_{N_{i}}$ of analytic
matrix functions, which converges on $K$ as $N_{i}\rightarrow \infty $ to a
solution of (\ref{MPm}). Repeating almost literally the argument, which
leads to Req. (2.2.8) of \cite{Pa-Sh:11}, we obtain%
\begin{equation}
\mathbf{E}\{G(z)\}=\left( (H_{S}-\Sigma _{S}f_{S}^{(N)}(z)\Sigma
_{S})\otimes \mathbf{1}_{R}+\mathbf{1}_{S}\otimes H_{R}-z\mathbf{1}_{S\cup
R}\right) ^{-1}+R_{N}(z),  \label{eqG}
\end{equation}%
where the $nN\times nN$ matrix $R_{N}(z)$ admits the bound%
\begin{equation}
||R_{N}(z)||\leq C_{S}(z)/N^{2}.  \label{remb}
\end{equation}%
Applying to (\ref{eqG}) -- (\ref{remb}) the formula%
\begin{equation*}
\mathrm{Tr}_{R}\varphi (A_{S}\otimes \mathbf{1}_{R}+\mathbf{1}_{S}\otimes
B_{R})=\sum_{l=1}^{N}\varphi (A_{S}+b_{l}),
\end{equation*}%
valid for any function $\varphi $, any $n\times n$ matrix $A_{S}$ and any $%
N\times N$ hermitian matrix $B_{R}$ with eigenvalues $\{b_{l}\}_{l=1}^{N}$
and using the definition (\ref{non}) of the density of states of the
reservoir, we obtain for the expectation (\ref{fN})%
\begin{equation}
f_{S}^{(N)}(z)=\int \frac{\nu _{N}(E)dE}{E+H_{S}-z-\Sigma
_{S}f_{S}^{(N)}(z)\Sigma _{S}}+r_{N}(z),  \label{MPmp}
\end{equation}%
where%
\begin{equation*}
||r_{N}(z)||\leq C_{S}(z)/N^{2}|\Im z|^{3}.
\end{equation*}%
It follows from (\ref{fN}) and (\ref{Gbou}) that
\begin{equation*}
||f_{S}^{(N)}(z)||\leq |\Im z|^{-1}.
\end{equation*}%
Hence, for any compact set $K\subset \mathbb{C}\diagdown \mathbb{R}$ there
exists a subsequence $\{f_{S}^{(N_{i})}\}_{N_{i}}$ of bounded analytic
matrix functions, which converges uniformly on $K$ to an analytic matrix
function $f$ and passing to the limit $N\rightarrow \infty $ in (\ref{MPmp})
we obtain (\ref{MPm}) for $z\in K$. The validity of the equation for any $%
z\in \mathbb{C}\diagdown \mathbb{R}$ follows from the analyticity of $f$ in $%
\mathbb{C}\diagdown \mathbb{R}$.

Let us show that (\ref{MPm}) is uniquely solvable in the class of analytic
matrix functions (\ref{nevm}). Choose
\begin{equation}
K\subset \{z\in \mathbb{C}:|\Im z|>||\Sigma _{S}||^{2}\}  \label{K}
\end{equation}%
and assume that there are two different solutions $f_{S}^{^{\prime }}$ and $%
f_{S}^{^{\prime }}$, i.e., $\max_{z\in K}||f_{S}^{(1)}(z)-f_{S}^{(2)}(z)||>0$%
. It follows then from (\ref{MPm}) and (\ref{nevm}) that
\begin{equation}
1\leq ||\Sigma _{S}||^{2}\int dE\nu (E)||(E-H_{S}-z-\Sigma
_{S}f_{S}^{^{\prime }}\Sigma _{S})^{-1}||\cdot ||(E-H_{S}-z-\Sigma
_{S}f_{S}^{^{\prime \prime }}\Sigma _{S})^{-1}||.  \label{MPmi}
\end{equation}%
In addition, (\ref{nevm}) implies the bounds $||(E-H_{S}-z-\Sigma
_{S}f_{S}^{^{\prime }}\Sigma _{S})^{-1}||\leq |\Im z|^{-1}$ and $%
||(E-H_{S}-z-\Sigma _{S}f_{S}^{^{\prime \prime }}\Sigma _{S})^{-1}||\leq
|\Im z|^{-1}$ (see (\ref{boure}) for details). This and (\ref{MPmi}) yield
the inequality $1\leq ||\Sigma _{S}||^{2}/|\Im z|^{2}$, which contradicts (%
\ref{K}).

The unique solvability of (\ref{MPm}) -- (\ref{nevm}) implies that the whole
sequence $\{f_{S}^{(N)}\}_{N}$ converges to the limit $f$, a unique solution
of (\ref{MPm}) -- (\ref{nevm}) everywhere in $\mathbb{C}\diagdown \mathbb{R}$%
.

It remains to prove that this fact together with (\ref{varab}) imply the
convergence with probability 1 of the sequence $\{g_{S}^{(N)}\}_{N}$ of
random analytic matrix functions to the solution of (\ref{MPm}) -- (\ref%
{nevm}).

It follows from the Tchebyshev inequality and (\ref{varab}) that for any $%
\delta >0$ and any non-real $z$
\begin{equation*}
\mathbf{P}\{||f_{S}^{(N)}(z)-g_{S}^{(N)}(z)||>\delta \}\leq \mathbf{Var}%
\{g_{S}^{(N)}(z)\}/\delta {^{2}}\leq {C}_{S}(z)/\delta {^{2}|\Im z|^{4}N^{2}}%
.
\end{equation*}%
Hence, for any non-real $z$ the series
\begin{equation*}
\sum_{N=1}^{\infty }\mathbf{P}\{||f_{S}^{(N)}(z)-g_{S}^{(N)}(z)||>\delta \}
\end{equation*}%
converges for any $\delta >0$, and by the Borel-Cantelli lemma and the
convergence of $\{f_{S}^{(N)}\}_{N}$ to $f_{S}$ we have with probability 1
for any non-real $z$
\begin{equation}
\lim_{N\rightarrow \infty }g_{S}^{(N)}(z)=f_{S}(z).  \label{lgnf}
\end{equation}%
Let us show that $g_{S}^{(N)}$ converges to $f_{S}$ uniformly on any compact
set of $\mathbb{C}\setminus \mathbb{R}$ with probability 1. Because of the
uniqueness of analytic continuation it suffices to prove that with the same
probability the limiting relation $\lim_{N\rightarrow \infty
}g_{S}^{(N)}(z_{j})=f_{S}(z_{j})$ is valid for all points of an infinite
sequence $\{z_{j}\}_{j\geq 1},\;\Im z_{j}\geq \eta _{0}>0$, possessing a
finite accumulation point. Denote $\Omega (z)$ the set of realizations,
where (\ref{lgnf}) is valid. According to the above $\mathbf{P}\{\Omega
(z_{j})\}=1,\,\forall j$. Hence,
\begin{equation*}
\mathbf{P}\Big\{\bigcap_{j\geq 1}\Omega (z_{j})\Big\}=1.
\end{equation*}%
This proves the uniform convergence of $g_{S}^{(N)}$ to $f_{S}$ on any
compact set of $\mathbb{C}\setminus \mathbb{R}$ with probability 1.

It is known that the one-to-one correspondence between the non negative
measures and the Stieltjes transforms is continuous in the topology of vague
convergence of measures and the uniform convergence on a compact set of $%
\mathbb{C}\setminus \mathbb{R}$ \cite{Pa-Sh:11}, Proposition 2.1.2. An
analogous assertion can be easily proved for the positive definite matrix
measures and their Stieltjes transforms. By using this fact and the uniform
convergence with probability 1 of $g_{S}^{(N)}$ to $f_{S}$ proved above, we
obtain (\ref{lehs}).

\section{Gibbs Distribution}

We will prove here our Result III, i.e., relation (\ref{limgi}), by using
Result II (see (\ref{MPm}) -- (\ref{nevm})) and our model (\ref{nuj}) of the
reservoir. Recall that we assume that the density $q$ of (\ref{nuj})
satisfies the conditions (i) and (ii) of Section 3, (\ref{phi}) -- (\ref%
{phidec}) in particular.

\medskip We will now use the matrix identity (see formulas (\ref{boure}) for
its justification)%
\begin{equation*}
-\frac{1}{\zeta }=i\int_{0}^{\infty }e^{it\zeta }dt,\;\Im \zeta >0,
\end{equation*}%
with $\zeta =z+\Sigma _{S}f_{S}(z)\Sigma _{S}-E-H_{S}$ to write the basic
equation (\ref{MPm}) with $\Im z>0$ as
\begin{equation}
f_{S}(z)=\int_{0}^{\infty }ie^{it(-\widetilde{H}+z)}\varphi ^{J}(t)dt,\;%
\widetilde{H}=H_{S}-\Sigma _{S}f_{S}(z)\Sigma _{S},  \label{FF}
\end{equation}%
where
\begin{equation}
\varphi (t)=\int e^{-i\varepsilon t}q(\varepsilon )d\varepsilon ,\;\varphi
(0)=1,\;|\varphi (t)|\leq 1,  \label{phi}
\end{equation}%
and we took into account that the Fourier transform of the convolution (\ref%
{nuj}) is $\varphi ^{J}$. Our proofs in this section are based on the above
integral representation of $f_{S}$.

Let us prove first that the limit
\begin{equation}
f_{S}(E):=\lim_{\delta \rightarrow 0^{+}}f_{S}(E+i\delta )  \label{fsd}
\end{equation}%
exists and is bounded for all real $E$ if $J$ is large enough. To this end
we first use the bounds%
\begin{equation}
||(M-\zeta \}^{-1}||\leq (\Im \zeta )^{-1},\;\Im \zeta >0;\;||e^{-itM}||\leq
1,\;t>0,  \label{boure}
\end{equation}%
valid for any complex matrix such that $\Im M>0$. Indeed, in this case $\Re
(-iM)\geq 0$. Matrices with this property are known as accretive and for
them the corresponding bounds $||(-iM+\zeta \}^{-1}||\leq (\Re \zeta
)^{-1},\;\Re \zeta >0;\;||e^{-tM}||\leq 1,\;t>0$ are valid (see e.g. \cite%
{Ka:76}, Section IX.1.6). Noting that $\Im f_{S}(z)>0,\;$ $\Im z>0$ implies $%
\Im \Sigma f_{S}(z)\Sigma >0$ for$\;\Im z>0$ and using the second bound of (%
\ref{boure}), we obtain from (\ref{FF})
\begin{equation}
||f_{S}(z)||\leq \int_{0}^{\infty }|\varphi (t)|^{J}dt,\;\Im z>0.
\label{Ffi}
\end{equation}%
It follows then from (\ref{phidec}) and the Hausdorff-Young inequality that
\begin{equation}
\int |\varphi (t)|^{J_{0}}dt\leq \left( \int q^{a}(\varepsilon )d\varepsilon
\right) ^{J_{0}/a} < \infty, \;J_{0}=a/(a-1)  \label{hay}
\end{equation}%
and we obtain from (\ref{phi}), (\ref{Ffi}) and (\ref{hay})
\begin{equation}
||f_{S}(z)||\leq \int_{0}^{\infty }|\varphi (t)|^{J_{0}}dt<\infty ,\;J\geq
J_{0}.  \label{nfbou}
\end{equation}%
It is important that the r.h.s. of the above bound is independent of $z$ for
$\Im z\geq 0$ and $J\geq J_{0}$, thus the same bound holds for $f_{S}(E)$ of
(\ref{fsd}) with any real $E$ and uniformly in $J\geq J_{0}$.

We will use now a stronger version of the inversion formula (\ref{SP})
corresponding to measures with a bounded density $m^{\prime }$:%
\begin{equation}
m^{\prime }(\lambda ):=\lim_{\Delta \rightarrow \{\lambda \}}m(\Delta
)/|\Delta |=\lim_{\delta \rightarrow 0^{+}}\pi ^{-1}\Im s(\lambda +i\delta ).
\label{FPd}
\end{equation}%
Since the same formula holds also in the matrix case, we will divide the
numerator and denominator of (\ref{rbe}) by $|\Delta |$ and pass to the
limit $\Delta \rightarrow \{E\}$ to obtain (\ref{rgg}).

In fact, an elaboration of the above argument allows us to proves the
relation%
\begin{equation}
\lim_{J\rightarrow \infty }||f_{S}(J\varepsilon )||=0,  \label{fvan}
\end{equation}%
valid for the limit (\ref{fsd}) with any $\varepsilon $ varying over a
finite interval which will be used below.

Indeed, it is shown in the Appendix that if $\varphi $ satisfies (\ref%
{phidec}), then for any $t_{0}>0$
\begin{equation}
\max_{t\geq t_{0}>0}|\varphi (t)|=\varphi _{0}<1.  \label{maxfi}
\end{equation}%
Choose a $\delta >0$ and write the integral in (\ref{Ffi}) as the sum of
integrals over $(0,\delta )$ and $(\delta ,\infty )$. Then it follows from (%
\ref{phi}), \ref{phidec}) and (\ref{hay}) that the first integral is bounded
by $\delta $ and the second is bounded by
\begin{equation}
\varphi _{0}^{J-J_{0}}\int_{\delta }^{\infty }|\varphi (t)|^{J_{0}}dt\leq
C\varphi _{0}^{J},\;J\geq J_{0},  \label{edec}
\end{equation}%
where $C$ is independent of $J$ and $\delta $. Thus, passing to the limit $%
J\rightarrow \infty $ and then $\delta \rightarrow 0$, we obtain (\ref{fvan}%
).

In fact, a bit more careful calculation yields that the r.h.s. of (\ref{fvan}%
) is of the order $J^{-1/2}$ if the first moment of $q$ is zero, and $J^{-1}$
if the first moment is not zero. This can be seen also in examples (\ref{ga0}%
) and (\ref{qd}).

Introduce the real and imaginary part of $f_{S}(J\varepsilon )$ (note that $%
f_{S}(J\varepsilon )$ is well defined in view of (\ref{Ffi})):%
\begin{equation}
f_{S}(J\varepsilon )=\mathcal{R}+i\mathcal{I}  \label{FCI}
\end{equation}%
and take into account that according to our assumptions the function $%
\varphi $ of (\ref{phi}) can be analytically continued into the lower
half-plane in $t$. This allows us to write the integral for $\mathcal{I}$,\
\ which determines $\gamma_J$ of (\ref{rgg}) according to (\ref{FPd}), as
the sum
\begin{equation}
\mathcal{I}:=\Im f_S(J\varepsilon)=I_{1}+I_{2}  \label{icii}
\end{equation}%
of integrals over $(0,-i\beta (\varepsilon ))$ and $(-i\tau \beta
(\varepsilon ),-i\beta (\varepsilon )+\infty )$, where $\beta (\varepsilon
)>0$ is the point where the function $h_{\varepsilon }$ of (\ref{sleg})
achieves its minimum, see (\ref{hpo}) -- (\ref{bepo}):

Changing $t$ to $-i\tau $ in $I_{1}$ we obtain%
\begin{equation}
I_{1}=\int_{0}^{\beta }e^{Jh_{\varepsilon }(\tau )}\mathcal{S}(\tau )d\tau
,\;  \label{I1}
\end{equation}%
where we write
\begin{equation}
e^{-\tau \widetilde{H}}=\mathcal{C}(\tau )+i\mathcal{S}(\tau ).  \label{ecs}
\end{equation}%
for $\widetilde{H}$ of (\ref{FF}) and $\beta $ instead of $\beta
(\varepsilon )$. Write also
\begin{equation}
\widetilde{H}=R+iI,\;R=H_{S}-\Sigma _{S}\mathcal{R}\Sigma _{S},\;I=\Sigma
_{S}\mathcal{I}\Sigma _{S}\geq 0.  \label{htri}
\end{equation}%
If $\widetilde{H}$ were a complex number but not a matrix (e.g., if $n=1$),
then
\begin{equation}
\mathcal{S}(\tau )=-e^{-\tau R}\sin \tau I=-e^{-\tau R}I\;c(\tau ),\;|c(\tau
)\leq \tau  \label{imeht}
\end{equation}%
and%
\begin{equation}
I_{1}=-\int_{0}^{\beta }e^{Jh_{\varepsilon }(\tau )}e^{-\tau R}I\;c(\tau
)d\tau .  \label{I1d}
\end{equation}%
This and the inequality (see (\ref{sleg}) -- (\ref{eeb}))
\begin{equation}
h(\tau )\geq (\varepsilon -\overline{\varepsilon })\tau ,  \label{lbouh}
\end{equation}%
would imply the bound
\begin{equation}
I_{1}\leq e^{\beta |R|}|I|\int_{0}^{\beta }\tau e^{-J|\varepsilon -\overline{%
\varepsilon }|\tau }d\tau \leq e^{\beta |R|}|\Sigma _{S}|^{2}\mathcal{I}%
/((\varepsilon -\overline{\varepsilon })J)^{2}.  \label{J2}
\end{equation}%
Comparing this with (\ref{icii}), we would conclude that the contribution of
$I_{1}$ in $\mathcal{I}$ of (\ref{icii}) is negligible as $J\rightarrow
\infty $, i.e.,%
\begin{equation}
\mathcal{I}=I_{2}\ (1+o(1)),\;J\rightarrow \infty .  \label{IJ2}
\end{equation}%
We are going now to find a matrix analog of the above asymptotic relation.

We note first that according to Appendix the matrix analog of (\ref{imeht})
is%
\begin{equation}
\mathcal{S}(\tau )=-\int_{0}^{\tau }e^{-(\tau -s)R}I\mathcal{C}(s)ds,\;||%
\mathcal{C}(s)||\leq se^{s||R||}\cosh (s||I||),  \label{duh}
\end{equation}%
hence the matrix analog of (\ref{I1d}) is in view of (\ref{htri})
\begin{align*}
I_{1}& =-\int_{0}^{\beta }e^{Jh_{\varepsilon }(\tau )}d\tau \int_{0}^{\tau
}e^{-(\tau -s)R}I\mathcal{C}(s)ds \\
& =\int_{0}^{\beta }e^{Jh_{\varepsilon }(\tau )}d\tau \int_{0}^{\tau
}e^{-(\tau -s)R}\Sigma _{S}\mathcal{I}\Sigma _{S}\mathcal{C}(s)ds.
\end{align*}%
It is convenient to view the r.h.s. of the above formula as the result of
application to the $n\times n$ matrix $\mathcal{I}$\ of the linear operator $%
\mathbf{A}$ acting in the space $\mathcal{M}_{n}(\mathbb{C})$ of complex $%
n\times n$ matrices, i.e., to write the formula as%
\begin{equation}
I_{1}=\mathbf{A}(\mathcal{I}).  \label{I1A}
\end{equation}%
Then we have from (\ref{duh}) just as in the scalar case (\ref{imeht}) -- (%
\ref{J2}):%
\begin{equation*}
||I_{1}||=||\mathbf{A}(\mathcal{I})||\leq e^{\beta ||R||}||\Sigma
_{S}||^{2}||\mathcal{I}\ ||\cosh (\beta ||\Sigma _{S}||^{2}||\mathcal{I}%
||)/((\varepsilon -\overline{\varepsilon })J)^{2}.
\end{equation*}%
By using the bounds $||R||\leq 2||H||$ and $||\mathcal{I}||\leq 1,$ which
following from (\ref{fvan}), we conclude that the matrix analog of bound (%
\ref{IJ2}) is also valid, i.e., the contribution of $I_{1}$ to the r.h.s. of
(\ref{icii}) is negligible as $J\rightarrow \infty $ in the matrix case as
well. This fact can be expressed via the operator $\mathbf{A}$:
\begin{equation}
||\mathbf{A}||\leq C_{S}/((\varepsilon -\overline{\varepsilon })J)^{2}.
\label{abou}
\end{equation}%
Consider now $I_{2}$ of (\ref{icii}). Changing the variable to $t=-i\tau
_{0}+\sigma $ and using $\mu _{J}$ of (\ref{df}), we obtain%
\begin{equation}
I_{2}/\mu _{J}(\varepsilon )=\sqrt{2\pi s^{\prime \prime }(\varepsilon )J}%
\Re e^{-\beta \widetilde{H}}\int_{0}^{\infty }e^{-i\sigma \widetilde{H}%
}e^{J\chi (\sigma )}d\sigma ,  \label{I2}
\end{equation}%
where%
\begin{equation}
\chi (\sigma )=i\varepsilon \tau +\log \psi _{\beta }(\sigma ),\;\psi
_{\beta }(\sigma )=\psi (\beta +i\sigma )/\psi (\beta ).  \label{chi}
\end{equation}%
We write the integral on the right of (\ref{I2}) as the sum of integrals
over $(0,\sigma _{0})$ and $(\sigma _{0},\infty )$, where $\sigma _{0}$ is
small enough:%
\begin{equation}
I_{2}=I_{21}+I_{22}.  \label{III}
\end{equation}%
In the first integral we use the expansion $\chi (\sigma )=2^{-1}s^{\prime
\prime }(\varepsilon )\sigma ^{2}(1+o(1)),\;\sigma \rightarrow 0$ of (\ref%
{chi}) yielding%
\begin{equation*}
I_{21}=\pi \Re e^{-\beta \widetilde{H}}(1+o(1)),\;J\rightarrow \infty ,
\end{equation*}%
and then (\ref{fvan}) and (\ref{htri}) imply%
\begin{equation}
I_{21}=\pi e^{-\beta H_{S}}(1+o(1)),\;J\rightarrow \infty .  \label{I21}
\end{equation}%
To deal with the second term
\begin{equation*}
I_{22}=\Re \int_{\sigma _{0}}^{\infty }e^{-(\tau _{0}+i\sigma )\widetilde{H}%
}e^{J\chi (\sigma )}d\sigma
\end{equation*}%
of (\ref{III}) we use first (\ref{fvan}) and (\ref{htri}) to obtain the
bound $||e^{-(\beta +i\sigma )\widetilde{H}}||\leq K(\varepsilon )$, where $%
K(\varepsilon )$ is independent of $J$. This and (\ref{I2}) -- (\ref{chi})
yield%
\begin{equation}
||I_{22}/\mu _{J}(\varepsilon )||\leq \sqrt{2\pi s^{\prime \prime
}(\varepsilon )J}K(\varepsilon )\int_{\delta }^{\infty }|\psi_{\varepsilon}(\sigma)|^J d\sigma ,  \label{i22}
\end{equation}%
where
\begin{equation*}
\psi _{\beta }(\sigma )=\psi (\beta +i\sigma )/\psi (\beta ).
\end{equation*}%
To estimate the integral on the right of (\ref{i22}) we will follow the
scheme of proof of (\ref{edec}), which was based on the bounds (\ref{maxfi})
and (\ref{phidec}). Thus, we have to prove the analogs of these bounds for $%
\psi _{\beta }$. To this end we use (\ref{psi}) and (\ref{qe}) to write%
\begin{equation}
\psi _{\beta }(\sigma )=\int e^{-i\varepsilon \sigma }Q(\varepsilon
)d\varepsilon .  \label{ppq}
\end{equation}%
We conclude that the $\psi _{\beta }$ is the Fourier transform of a
non-negative function of unit integral, thus it satisfies (\ref{phi}), i.e.,
$|\psi _{\beta }(\sigma )|\leq 1,\;\psi _{\beta }(0)=1$. It follows then
from Appendix that in this case we have an analog of (\ref{maxfi}) as well:
\begin{equation}
\max_{\sigma \geq \sigma _{0}>0}\left\vert \psi _{\beta }(\sigma
)\right\vert :=\kappa _{1}<1.  \label{mpp}
\end{equation}%
Furthermore, according to (\ref{phi}) and (\ref{ppq}), an analog of (\ref%
{phidec}) for $\psi _{\beta }$ is
\begin{equation*}
\int Q^{a}(\varepsilon )d\varepsilon =\int e^{-a\beta \varepsilon
}q^{a}(\varepsilon )d\varepsilon <\infty .
\end{equation*}%
It is easy to see this is indeed true in view of conditions (i) and (ii) of
Section 3, (\ref{phidec}) in particular.

This, (\ref{III}) and (\ref{I21}) yield $I_{2}=\pi e^{-\beta
H_{S}}(1+o(1)),\;J\rightarrow \infty $. Now use (\ref{I1A}) to write (\ref%
{icii}) as $\mathcal{I=}\mathbf{A}\mathcal{(I)}+I_{2}$. Then (\ref{abou})
and (\ref{I2}) imply
\begin{equation*}
\mathcal{I}/\mu _{J}(\varepsilon )=(\mathbf{1}-\mathbf{A)}^{-1}I_{2}/\mu
_{J}(\varepsilon )=\pi e^{-\beta H_{S}}(1+o(1)),\;J\rightarrow \infty .
\end{equation*}%
This and (\ref{FPd}) yield (\ref{gamu}). It remains then to divide the
numerator and the denominator in (\ref{rgg}) by $\mu _{j}$ and to use (\ref%
{gamu}) to obtain (\ref{limgi}).

\medskip \textbf{\large Acknowledgement}. We thank Chris Jarzynski for
helpful comments. We also thank the Institute for Advanced Study in Princeton
N.J. for hospitality when the part this work was done. Supported in part by NSF Grant
DMR-1104501.

\section*{Appendix}

(i) \textit{Proof of (\ref{maxfi}).} We prove here that if $q$ is
non-negative, continuous and of unit integral, then we have (\ref{maxfi}) in
addition to ( \ref{phi}).

Indeed, the equality $\varphi (t^{\prime })=0\,\ $ for some $t^{\prime }\neq
0$ implies the equality%
\begin{equation*}
\int (1-\cos \varepsilon t^{\prime })q(\varepsilon)d\varepsilon=0,
\end{equation*}%
which is impossible for any non-negative and non-zero $q$ satisfying (\ref{phidec}).
This implies (\ref{maxfi}).

\bigskip (ii) \textit{Proof of (\ref{duh}).} The Duhamel formula%
\begin{equation*}
e^{A+B}=e^{A}+\int_{0}^{1}e^{(1-s)A}Be^{s(A+B)}ds,
\end{equation*}%
valid for any two (generally non-commuting) matrices, yields the formulas
for the terms of the r.h.s. of (\ref{ecs})%
\begin{equation*}
\mathcal{S}(\tau )=-\int_{0}^{\tau }e^{-(\tau -s)R}I\mathcal{C}(s)ds,
\end{equation*}%
and%
\begin{equation*}
C(\tau )=e^{-\tau R}+\int_{0}^{\tau }e^{-(\tau -s)R}I\mathcal{S}(s)ds
\end{equation*}%
hence%
\begin{equation*}
\mathcal{C}(s)=e^{-sR}+\sum_{l=1}^{\infty
}(-1)^{l}\int_{0}^{s}e^{(s-s_{1})R}Ids_{1}%
\int_{0}^{s_{1}}e^{-(s_{1}-s_{2})R}Ids_{2}...%
\int_{0}^{s_{2l-1}}e^{-(s_{2l-1}-s_{2l})R}Ie^{-s_{2l}R}ds_{2l}.
\end{equation*}%
These formulas combined with the standard upper bounds for the terms of the
series lead to (\ref{duh}).

In the commutative case the formulas are%
\begin{equation*}
e^{-\tau R}\sin \tau I=e^{-\tau R}Ic(\tau),\;c(\tau)=\int_{0}^{\tau}\cos
sI\;ds,
\end{equation*}%
i.e., coincide with (\ref{imeht})). The non-commutative analog (\ref{duh})
of (\ref{imeht}) is more rough, since we do not take into account the
alternating signs in the above series, hence the effects of strong
cancelations of its terms.

\end{document}